\documentclass[letterpaper, 10 pt, conference]{ieeeconf}

\IEEEoverridecommandlockouts                              
\usepackage{graphicx}
\usepackage{ifpdf}
\usepackage{amssymb}

\usepackage{colortbl}

\usepackage{array}
\usepackage{multirow}

\def\theequation{\thesection.\arabic{equation}}

\newcommand{\be}{\begin{equation}}
\newcommand{\ee}{\end{equation}}
\newcommand{\bd}{\begin{displaymath}}
\newcommand{\ed}{\end{displaymath}}
\newcommand{\bea}{\begin{eqnarray}}
\newcommand{\eea}{\end{eqnarray}}

\newcommand{\Hi} {{\cal H}^\infty}

\newcommand{\compleib}{COMPL$_e$IB}

\def\matlab{{\sc \mbox{matlab}}}
\def\hifoo{{\sc \mbox{hifoo}}}
\def\hanso{{\sc \mbox{hanso}}}
\def\mosek{{\sc \mbox{mosek}}}
\def\slicot{{\sc \mbox{slicot}}}

\title{\LARGE $\Hi$ Strong Stabilization via HIFOO,\\ a Package for
Fixed-Order Controller Design}

\author{Suat Gumussoy, Marc Millstone and Michael L. Overton
\thanks{Suat Gumussoy is with the Computer Science Department, Katholieke Universiteit Leuven, Belgium,
 {\tt\small gumussoysuat@yahoo.com}}
\thanks{Marc Millstone and Michael L. Overton are with the Courant Institute of Mathematical
Sciences, New York University, 251 Mercer St., New York, NY 10012,
{\tt\small millstone@cs.nyu.edu},
{\tt\small overton@cs.nyu.edu}}%
}

\date{}

\begin{document}
\maketitle

\begin{abstract}
We report on our experience with strong stabilization using \hifoo,
a toolbox for $\Hi$ fixed-order controller design. We applied
\hifoo\ to $21$ fixed-order stable $\Hi$ controller design problems
in the literature, comparing the results with those published for
other methods. The results show that \hifoo\ often achieves good
$\Hi$ performance with \emph{low-order stable} controllers, unlike
other methods in the literature.\end{abstract}

\section{Introduction}
\label{intro} By \emph{$\Hi$ strong stabilization} we mean the
following: given a linear time-invariant (LTI)
multi-input-multi-output (MIMO) system, we are interested in
designing a controller that stabilizes the system in closed-loop,
reducing the $\Hi$ norm of its closed loop transfer function as much
as possible, with the additional constraint that the
\emph{controller is stable}. In addition, we require the controller
to have a fixed order, specified by the designer. Stable controllers
offer several advantages, specifically with respect to disturbance
rejection, tracking and modeling uncertainties \cite{ZDG-ROC-96}.
They offer protection against sensor failures and actuator
saturation \cite{V-CSS-85}. Furthermore, low-order controllers are
simpler and therefore easier to implement than full-order
controllers, whose order equals the order of the plant, and may
therefore offer more confidence for practical use.

Optimal and parameterized suboptimal full-order $\Hi$ controllers
for LTI MIMO systems can be designed by well-known methods in the
literature \cite{DGKF, RobustControlToolbox}. However, the practical
value of controllers obtained by these methods is limited by the
fact that they are full-order and are not generally stable.

There are various methods in the control literature to design stable
$\Hi$ controllers \cite{CZ-TAC-01, CWL-CDC-03, CC-TAC-01,
HinfssMIMO_GO, LS-SCL-02, ZO-TAC-99, ZerenOzbayAutomatica}. All the
controllers obtained by these methods have order greater than or
equal to the plant order, specifically
\begin{itemize}
  \item the plant order \cite{CC-TAC-01,LS-SCL-02};
  \item the plant order plus a free parameter $Q$ \cite{CZ-TAC-03};
  \item double the plant order \cite{ZO-TAC-99, ZerenOzbayAutomatica, HinfssMIMO_GO};
  \item double the plant order plus the order of a weighting function
\cite{CZ-TAC-01};
  \item three times the plant order \cite{CWL-CDC-03}.
\end{itemize}
Thus, these controllers are not practical for high-order plants.

It seems that the reason there is not much literature on designing
low-order stable $\Hi$ controllers is that the order of stable
controllers could be very large when the plant pole-zero locations
in the right-half-plane are close to violating the parity interlacing
property
According to \cite{CLC-IET-07}, it is for this reason that
\textit{instead of pursuing 'minimal-order'
stable controllers, researchers focus on providing alternative
methods to solve the problems}, usually resulting in high-order
controllers.

In this paper, we report on our experience applying the \hifoo\
toolbox (Version 1.75) to fixed-order strong stabilization $\Hi$
controller design problems, attempting to minimize the $\Hi$ norm of
the transfer function for the closed loop plant using a stable
controller. This is a difficult optimization problem due to the
non\-convexity and non\-smoothness of the objective function and the
stability constraint. \hifoo\ 1.75 uses a hybrid algorithm for
nonsmooth, nonconvex optimization based on several techniques to
attempt to find fixed-order stable controllers achieving minimal
closed-loop $\Hi$ norm. \hifoo\ 1.0 was originally presented in
\cite{BHLO-IFAC-06}, but the original version did not support strong
stabilization. \hifoo\ does not have any restrictions on plant or
controller such as nullity or full-rank conditions. It allows the
controller order to be specified by the user, unlike other methods
in the literature.

\hifoo\ is freely available \matlab\ code$^1$
\footnotetext[1]{http://www.cs.nyu.edu/overton/software/hifoo/}
and has been designed to be easy to use. It is built on the \hanso\
optimization package, freely available at the same website. It does
not require any external software beyond the \matlab\ Control System
Toolbox, but it runs much faster if the {\tt linorm} function of the
\slicot\ package is installed and in the \matlab\ path (available
commercially from {\tt www.slicot.de}, but freely available from the
\hifoo\ webpage for noncommercial use with \hifoo\ using \matlab\
running under Windows).  \hifoo\ also makes use of the {\tt
quadprog} quadratic programming solver from \mosek\ or the \matlab\
Optimization Toolbox if it is installed and in the \matlab\ path,
but this is not required. Our experiments used \matlab\ 2006a with
{\tt linorm} and {\tt quadprog} installed.

We applied \hifoo\ to various benchmark plants in the literature and
compared our results with published results based on other
techniques. Our experience is that \hifoo\ gives very good
experimental results for large sets of data. In particular, we find
that it is often possible to obtain stable $\Hi$ controllers
achieving small closed-loop $\Hi$ norm even when the order of the
controller is fixed to be much less than the order of the plant.

The rest of the paper is organized as follows. The problem of
fixed-order strongly stable $\Hi$ controller design is described and
the optimization method used by \hifoo\ is summarized in Section
\ref{problemandopt}. The benchmark plants are specifed in Section
\ref{benchmarkproblems}. Our computational results and comparisons
with those published for other methods are given in Section
\ref{examples}.
Concluding remarks are in Section \ref{concluding}.

\section{Problem Formulation and Optimization Method} \label{problemandopt}

The state-space equations of a generalized plant $G$ are
\bea \label{eq:GeneralizedPlant}
\nonumber \dot{x}(t)&=&Ax(t)+B_1w(t)+B_2u(t), \\
\nonumber z(t)&=&C_1x(t)+D_{11}w(t)+D_{12}u(t), \\
y(t)&=&C_2x(t)+D_{21}w(t)+D_{22}u(t)
\eea and the state-space realization for the controller $K$ is
\bea \label{eq:Controller}
\nonumber \dot{x}_K(t)&=&A_Kx_K(t)+B_Ky(t), \\
u(t)&=&C_Kx_K(t)+D_Ky(t),
\eea where $A\in\mathcal{R}^{n\times n}$,
$D_{12}\in\mathcal{R}^{p_1\times m_2}$,
$D_{21}\in\mathcal{R}^{p_2\times m_1}$, with other matrices having
compatible dimensions, and $A_K\in\mathcal{R}^{n_K\times n_K}$, with
$B_K, C_K, D_K$ having dimensions that are compatible with $A_K$ and
the generalized plant matrices. The controller order $n_K$ is fixed,
so it can be specified by the designer.

The signals $(z,w,y,u)$ respectively represent the regulated outputs,
the exogenous inputs (including disturbance and commands), the
measured (or sensor) inputs, and the control inputs. The transfer function from the input
$w$ to output $z$ is denoted $T_{zw}$; see \cite{DGKF} for details.
The optimal $\Hi$ controller design can be formulated as
minimization of the closed loop $\Hi$ norm function \bea
 \inf_{K~\textrm{stabilizing}}
\|T_{zw}\|_\infty  , \eea where the constraint specifies that
$K$ internally stabilizes the closed-loop system.

In this paper, we impose the additional constraint that the
\emph{controller is stable}, so that we wish to minimize
\bea
\inf_{K~\textrm{stabilizing}~\textrm{and}
~K~\textrm{stable}}
\|T_{zw}\|_\infty.
\eea
Let us use $\alpha(X)$ to denote the spectral abscissa of a matrix
$X$, i.e., the largest of the real parts of the eigenvalues.
Thus, not only do we require that $\alpha(A_{CL}) < 0$, where
$A_{CL}$ is the closed-loop system matrix,
but we also require that $\alpha(A_K) < 0$.
The feasible set for $A_K$, that is the set of stable matrices, is
not a convex set and has a boundary that is not smooth. It has been
studied extensively, see e.g.  \cite{BurkeOverton01,HinrPrit05}.

As with previous versions \cite{BHLO-IFAC-06,Millstone2006}
\hifoo\ uses two phases: stabilization
and performance optimization.  In the stabilization phase, \hifoo\ 1.75
proceeds to minimize $\max(\alpha(A_{CL},\epsilon \alpha(A_K))$,
where $\epsilon$ is a positive parameter that will be described shortly,
until it finds a controller $K$ for which this quantity is negative, i.e.,
the controller is stable \emph{and} stabilizes the closed loop system.
If it cannot find such a controller, \hifoo\ will
return with a message to that effect.  In the performance optimization
phase, \hifoo\ 1.75 looks for a local minimizer of
\bea
 f(K) = \left \{ \begin{array}{c}\infty
~\mathrm{if}~\max(\alpha(A_{CL}),\alpha(A_K)) \geq 0 \\
\max ( \|T_{zw}\|_{\infty} ,\epsilon \|K\|_\infty )
 \mathrm{~otherwise},\\
\end{array} \right .
\eea
where
\bea
\|K\|_\infty = \sup_{\Re s = 0} \|C_K (s I - A_K)^{-1} B_K + D_K\|_2.
\eea
The motivation for the introduction of $\epsilon$ is that the
principal design goal is to stabilize the closed loop system and
minimize $\|T_{zw}\|_{\infty}$, indicating that
$\epsilon$ should be relatively small, but the $\epsilon \|K\|_\infty$
term prevents the $\Hi$ norm of the controller from
growing too large, which the stability constraint by itself will not.
Because of the stabilization phase, the performance
optimization phase begins with
a finite value for $f(K)$.  When it subsequently
encounters an instance of $K$ for which $f(K)=\infty$ it is
rejected by the line search which insists on a reduction in the
objective at every iteration.

The optimization code called by \hifoo\ in both phases is \hanso,
which implements a hybrid algorithm for nonsmooth, nonconvex
optimization, based on the following elements: a quasi-Newton
algorithm (BFGS) provides a fast way to approximate a local
minimizer; a local bundle method attempts to verify local optimality
for the best point found by BFGS, and if this does not succeed,
gradient sampling \cite{BurkeSIAM05,BurkeTAC06} attempts to refine
the approximation of the local minimizer, returning a rough
optimality measure. The local bundle and gradient sampling methods
are not invoked if the quadratic programming code {\tt quadprog} is
not in the \matlab\ path.  All three of these
optimization techniques use gradients which are automatically
computed by \hifoo. No effort is made to identify the exceptional
points where the gradients fail to exist. The algorithms are not
defeated by the discontinuities in the gradients at exceptional
points. The BFGS phase builds a highly ill-conditioned Hessian
approximation matrix, and the bundle and gradient sampling final
phases search for a point in parameter space for which a convex
combination of gradients at nearby points has small norm. More
details are given in \cite{BHLO-IFAC-06}.

Because \hifoo\ uses randomized starting
points, and also the gradient sampling phase involves randomization,
the same results are not obtained every time \hifoo\ is run.
In the results reported below, we made multiple runs setting
$\epsilon$ to $10^{-2}$, $10^{-3}$,$10^{-4}$,$10^{-5}$, and $10^{-6}$,
and running each case 10 times.  Each result in the tables in
Section \ref{examples} reports the lowest value for $\|T_{zw}\|_{\infty}$
obtained over all these runs.  We did not attempt to compare the running
times of different methods. In our view, one of the biggest advantages of
\hifoo\ is its ease of use. Generally, the running time requirements
for computation of controllers are not nearly as important as the
performance and safety aspects of the computed controllers.
Implementing any controller is far more work than computing it, so
the key aspect of running time in computing a controller is that it
should not be longer than the designer is willing to wait.  For this
reason \hifoo\ accepts an option, {\tt options.cpumax}, which
controls the running time.  Better performance may be obtained if a
larger value of {\tt options.cpumax} is specified. We set {\tt
options.cpumax} to 300 (5 minutes) in all of our tests.

\section{Benchmark Problems} \label{benchmarkproblems}

Benchmark examples for stable $\Hi$ controller design were chosen
from both applied and academic test problems, as follows.
\begin{enumerate}
  \item \textbf{Zeren-\"Ozbay Example:} A $5^{\textrm{th}}$-order plant given in \cite{ZerenOzbayAutomatica}.
 For this example, the optimal full-order $\Hi$
 controller is unstable and furthermore the central
 controller \cite{DGKF} for any closed-loop $\Hi$ norm is unstable.

  \item \textbf{Cao-Lam Example:} A $2^{\textrm{nd}}$-order
  plant given in \cite{CL-AUT-00}.
  \item \textbf{Choi-Chung Example:} A $4^{\textrm{th}}$-order
  plant given in \cite{CC-TAC-01}.
  \item \textbf{Four-Disk System:} An $8^{\textrm{th}}$-order four-disk system with
  noncolocated sensors and actuators given in \cite{FourDiskSystem,
  CZ-TAC-01}.
  \item \textbf{AC$8$:} A $9^{\textrm{th}}$-order state-space model of the linearized vertical plane dynamics of an aircraft \cite{TransportPlane};
  \item \textbf{HE$1$:} A $4^{\textrm{th}}$-order model of the longitudinal motion of a
  VTOL helicopter \cite{VTOL};
  \item \textbf{REA$2$:} A $4^{\textrm{th}}$-order chemical reactor model
  \cite{Chemicalreactor};
  \item \textbf{AC$10$:} A $55^{\textrm{th}}$-order aeroelastic model
  of a modified Boeing B-$767$ airplane \cite{AC10example};
  \item \textbf{BDT$2$:} An $82^{\textrm{nd}}$-order realistic model of a binary distillation
  tower \cite{BDTexample};
  \item \textbf{HF$1$:} A $130^{\textrm{th}}$-order one-dimensional model for heat flow
  in a thin rod \cite{HFexample};
  \item \textbf{CM$4$:} A $240^{\textrm{th}}$-order cable
  mass model for nonlinear dynamic response of a relief valve protecting a
  pneumatic system from overpressure \cite{CMexample};
  \item \textbf{PA:} A $5^{\textrm{th}}$-order model of a
  piezoelectric bimorph actuator system \cite{Piezoelectric};
  \item \textbf{HIMAT:} A $20^{\textrm{th}}$-order model of
  an experimental highly maneuverable (HIMAT) airplane \cite{HIMATExGod};
  \item \textbf{VSC:} A $4^{\textrm{th}}$-order quarter-car model representing
   characteristics of a real suspension system \cite{VSC1997};
  \item \textbf{AUV:} $3^{\textrm{rd}}$,
  $5^{\textrm{th}}$ and $6^{\textrm{th}}$-order linearized models of an autonomous underwater
  vehicle for speed, heading and depth autopilots respectively \cite{AUVEx};
  \item \textbf{Enns' Example:} An $8^{\textrm{th}}$-order plant used as an academic test problem for
  designing reduced-order $\Hi$ controllers \cite{Enns_example};
 \item \textbf{Wang's Example:} A $4^{\textrm{th}}$-order plant used as a theoretical benchmark problem for
  designing reduced-order $\Hi$ controllers \cite{Wang2003}, Example~$6.2$.
\end{enumerate}

Examples $1-4$ are collected from various papers specifically
concerned with strongly stable $\Hi$ controller design. The plants
$5-17$ were collected in \cite{HIFOO_08} as benchmark examples for
fixed-order $\Hi$ controller design without any stability constraint
on the controller. Examples $5-15$ are taken from real applications
and $16-17$ are academic test problems. The problem data for
examples $5-12$ are obtained from the \compleib $\;$library
\cite{Compleib} and those for examples $13-17$ are collected from
various papers in the literature. In the runs reported in the next
section, the strong stabilization constraint is imposed for all
examples. We do not give running times in this paper, but times for
the results reported in \cite{HIFOO_08} are available on the
web.$^2$ \footnotetext[2]{
http://www.cs.nyu.edu/overton/papers/pdffiles/acc08times.pdf}

\section{Results on Benchmark Problems} \label{examples}

In Tables I-VI, we compare the performance of \hifoo\ with other
methods from the strong stabilization literature on examples 1-4.
Tables VII-VIII show results obtained by \hifoo\ when the strong
stabilization constraint is imposed on examples 5-17 (there are no
results from the literature to compare for these examples).  In all
the tables, the controller order is shown by $n_K$, and
$\gamma_{n_K}$ shows the $\Hi$ performance achieved for this order
using the method indicated.  In Tables I-VI, the lines in the table
shaded in \emph{light gray} show results for the various strong
stabilization methods in the literature, which all produce
controllers with order greater than the order of the plant, as
mentioned in Section \ref{intro}. In all the tables, the line shaded
in \emph{dark gray} (labeled {\bf full} in Tables I-VI),  shows,
for $n_K$, the order of the plant and, for $\gamma_{n_K}$, the $\Hi$
performance for the optimal full-order controller computed using the
\verb"hinfsyn" routine of \cite{RobustControlToolbox} (see also
\cite{DGKF}). The $\Hi$ performance of the full-order controller is
a lower bound for the achievable $\Hi$ norm by any order controller
and is therefore a measure of performance for all methods. The
\emph{unshaded} lines below the full-order controller line show the
results obtained by \hifoo\ for various specified controller orders.
The last column in Tables I-VI indicates whether the
controller is stable.

\subsection{Zeren-\"Ozbay Example}

\begin{table}[h]
\caption{Comparison on Zeren-\"Ozbay Example}
\label{zerenozbaytable} \vspace{-6mm}
\begin{center}
\begin{tabular}{cccc}
   &  &  &  \\
  \hline
   &  &  &  \\
  $n_K$& $\gamma_{n_K}$   & Methods & Controller\\
   & & & Stability \\
  \hline
 \rowcolor[gray]{.9}  &  &  &  \\
 \rowcolor[gray]{.9} $10$ &  $42.51$ & \cite{ZerenOzbayAutomatica} & Stable \\
 \rowcolor[gray]{.9} $10$ &  $35.29$ & \cite{HinfssMIMO_GO} & Stable\\
 \rowcolor[gray]{.9} $6$  &  $34.44$ & \cite{CZ-TAC-03} & Stable   \\
 \rowcolor[gray]{.8} $5$  &  $34.24$ & {\bf full} &  Unstable \\
 $5$  &  $34.81$ & \hifoo & Stable\\
  $4$  &  $34.97$ & \hifoo & Stable\\
  $3$  &  $34.94$ & \hifoo & Stable\\
  $2$  &  $41.16$ & \hifoo & Stable\\
  $1$  &  $57.32$ & \hifoo & Stable\\
  \hline
\end{tabular}
\end{center}
\end{table}

Results for this example are given in Table \ref{zerenozbaytable}.
The plant order is 5 and the optimal full-order controller is
unstable. The performance of the method \cite{CZ-TAC-03} is good as
it finds a stable $6^{\textrm{th}}$ order controller with a
closed-loop $\Hi$ norm close to the optimal full-order performance.
However, \hifoo\ finds a stable $3^{\textrm{rd}}$ order controller
with nearly the same $\Hi$ norm.

\subsection{Cao-Lam Example}

\begin{table}[h]
\caption{Comparison on Cao-Lam Example} \label{caojamestable}
\vspace{-6mm}
\begin{center}
\begin{tabular}{cccc}
   &  &  &  \\
  \hline
   &  &  &  \\
  $n_K$& $\gamma_{n_K}$  & Methods & Controller\\
   & & & Stability \\
  \hline
 \rowcolor[gray]{.9}  &  &  &  \\
 \rowcolor[gray]{.9} $8$ & $1.29338$ & \cite{CLC-IET-07} & Stable \\
 \rowcolor[gray]{.9} $4$ & $1.36994$ & \cite{LS-SCL-02} & Stable   \\
 \rowcolor[gray]{.9} $4$ & $1.36957$ & \cite{HinfssMIMO_GO} &  Stable \\
 \rowcolor[gray]{.9} $4$ & $1.36814$ & \cite{CLC-IET-07} & Stable\\
 \rowcolor[gray]{.8} $2$ & $1.29022$ & {\bf full} & Unstable\\
  $2$  &  $1.36957$ & \hifoo & Stable\\
  $1$  &  $1.36957$ & \hifoo & Stable\\
  \hline
\end{tabular}
\end{center}
\end{table}

Results for this example are given in Table \ref{caojamestable}. The
plant order is 2 and, as in the previous example, the optimal
full-order $\Hi$ controller is unstable.  The method of
\cite{CLC-IET-07} finds a stable controller with nearly the same
$\Hi$ performance, but it uses an $8^\textrm{th}$ order controller.
\hifoo\ finds a stable $1^\textrm{st}$ order controller with less
than 10\% increase in $\Hi$ performance, approximately the same as
that found by the other methods for $4^\textrm{th}$ order
controllers.

\subsection{Choi-Chung Example}

\begin{table}[h]
\caption{Comparison on Choi-Chung Example} \label{choichungtable}
\vspace{-6mm}
\begin{center}
\begin{tabular}{cccc}
   &  &  &  \\
  \hline
   &  &  &  \\
  $n_K$& $\gamma_{n_K}$   & Methods & Controller\\
   & & & Stability \\
  \hline
 \rowcolor[gray]{.9}  &  &  &  \\
 \rowcolor[gray]{.9} $16$ & $25.430$ & \cite{CLC-IET-07} & Stable \\
 \rowcolor[gray]{.9} $12$ & $21.787$ & \cite{CWL-CDC-03} & Stable   \\
 \rowcolor[gray]{.9} $8$ & $43.167$ & \cite{CC-TAC-01} &  Stable \\
 \rowcolor[gray]{.9} $8$ & $37.551$ & \cite{ZO-TAC-99} & Stable\\
 \rowcolor[gray]{.9} $8$ & $32.557$ & \cite{HinfssMIMO_GO} & Stable\\
 \rowcolor[gray]{.9} $8$ & $24.790$ & \cite{CLC-IET-07} & Stable\\
 \rowcolor[gray]{.8} $4$ & $12.015$ & {\bf full} & Unstable\\
  $4$  &  $16.612$ & \hifoo & Stable\\
  $3$  &  $16.486$ & \hifoo & Stable\\
  $2$  &  $20.797$ & \hifoo & Stable\\
  $1$  &  $62.638$ & \hifoo & Stable\\
  \hline
\end{tabular}
\end{center}
\end{table}

Results for this example are given in Table
\ref{choichungtable}. We see again that the stable $\Hi$ controller
design methods in the literature are conservative in terms of
controller order. \hifoo\ achieves better $\Hi$ performance than the
other methods with a lower-order stable controller. 

\subsection{Four-Disk System}
\begin{table}[h]
\caption{Comparison on Four-Disk System, $\beta=10^{-1}$}
\label{fourdisksystem_1table} \vspace{-6mm}
\begin{center}
\begin{tabular}{cccc}
   &  &  &  \\
  \hline
   &  &  &  \\
  $n_K$& $\gamma_{n_K}$   & Methods & Controller\\
   & & & Stability \\
  \hline
 \rowcolor[gray]{.9}  &  &  &  \\
 \rowcolor[gray]{.9} $24$ & $0.237$ & \cite{CZ-TAC-01} & Stable \\
 \rowcolor[gray]{.9} $16$ & $0.245$ & \cite{ZO-TAC-99} & Stable   \\
 \rowcolor[gray]{.9} $16$ & $0.241$ & \cite{HinfssMIMO_GO} & Stable \\
 \rowcolor[gray]{.8} $8$ & $0.232$ & {\bf full} & Unstable\\
  $8$ & $0.235$ & \hifoo & Stable\\
  $7$ & $0.236$ & \hifoo & Stable\\
  $6$ & $0.236$ & \hifoo & Stable\\
  $5$ & $0.235$ & \hifoo & Stable\\
  $4$ & $0.274$ & \hifoo & Stable\\
  $3$ & $0.307$ & \hifoo & Stable\\
  $2$ & $0.347$ & \hifoo & Stable\\
  $1$ & $0.649$ & \hifoo & Stable\\
  \hline
\end{tabular}
\end{center}
\end{table}

\begin{table}[h]
\caption{Comparison on Four-Disk System, $\beta=10^{-2}$}
\label{fourdisksystem_2table} \vspace{-6mm}
\begin{center}
\begin{tabular}{cccc}
   &  &  &  \\
  \hline
   &  &  &  \\
  $n_K$& $\gamma_{n_K}$   & Methods & Controller\\
   & & & Stability \\
  \hline
 \rowcolor[gray]{.9}  &  &  &  \\
 \rowcolor[gray]{.9} $24$ & $0.151$ & \cite{CZ-TAC-01} & Stable \\
 \rowcolor[gray]{.9} $16$ & $0.178$ & \cite{ZO-TAC-99} & Stable   \\
 \rowcolor[gray]{.9} $16$ & $0.176$ & \cite{HinfssMIMO_GO} &  Stable \\
 \rowcolor[gray]{.8} $8$ & $0.141$ & {\bf full} & Unstable\\
  $8$ & $0.152$ & \hifoo & Stable\\
  $7$ & $0.153$ & \hifoo & Stable\\
  $6$ & $0.153$ & \hifoo & Stable\\
  $5$ & $0.152$ & \hifoo & Stable\\
  $4$ & $0.212$ & \hifoo & Stable\\
  $3$ & $0.276$ & \hifoo & Stable\\
  $2$ & $0.314$ & \hifoo & Stable\\
  $1$ & $0.634$ & \hifoo & Stable\\
  \hline
\end{tabular}
\end{center}
\end{table}

\begin{table}[h]
\caption{Comparison on Four-Disk System, $\beta=10^{-3}$}
\label{fourdisksystem_3table} \vspace{-6mm}
\begin{center}
\begin{tabular}{cccc}
   &  &  &  \\
  \hline
   &  &  &  \\
  $n_K$& $\gamma_{n_K}$   & Methods & Controller\\
   & & & Stability \\
  \hline
 \rowcolor[gray]{.9}  &  &  &  \\
 \rowcolor[gray]{.9} $24$ & $0.132$ & \cite{CZ-TAC-01} & Stable \\
 \rowcolor[gray]{.9} $16$ & $0.170$ & \cite{ZO-TAC-99} & Stable   \\
 \rowcolor[gray]{.9} $16$ & $0.170$ & \cite{HinfssMIMO_GO} &  Stable \\
 \rowcolor[gray]{.8} $8$ & $0.122$ & {\bf full} & Unstable\\
  $8$ & $0.142$ & \hifoo & Stable\\
  $7$ & $0.143$ & \hifoo & Stable\\
  $6$ & $0.145$ & \hifoo & Stable\\
  $5$ & $0.154$ & \hifoo & Stable\\
  $4$ & $0.208$ & \hifoo & Stable\\
  $3$ & $0.274$ & \hifoo & Stable\\
  $2$ & $0.314$ & \hifoo & Stable\\
  $1$ & $0.634$ & \hifoo & Stable\\
  \hline
\end{tabular}
\end{center}
\end{table}

The results for the Four-Disk System with three different parameter
values are shown in Tables IV-VI. One can see that the design
objectives (stability and low closed-loop $\Hi$ norm) are achieved
by \hifoo\ using low-order controllers. \hifoo\ achieves the same
$\Hi$ performance as the other methods with a $5^{\textrm{th}}$
order stable controller whereas the controllers obtained by the
other methods have order $16, 16$ and $24$ respectively.


\subsection{Other Benchmark Examples}
\begin{table}[h!]
\caption{Results of \hifoo\ on Other Benchmark Examples (Low-Medium
Order Plants). The first line shows the plant order and full-order
controller performance, with an asterisk indicating an unstable
full-order controller.} \label{otherexamples1_table} \vspace{-5mm}
\begin{center}
\begin{tabular}{>{\columncolor[rgb]{0.9,0.9,0.9}}c>{\columncolor[rgb]{0.9,0.9,0.9}}c
   cc
   >{\columncolor[rgb]{0.9,0.9,0.9}}c>{\columncolor[rgb]{0.9,0.9,0.9}}c
   cc}
   \hline
   \multicolumn{8}{c}{} \\
   \multicolumn{2}{>{\columncolor[rgb]{0.9,0.9,0.9}}c}{HE$1$} & \multicolumn{2}{c}{REA$2$}
   & \multicolumn{2}{>{\columncolor[rgb]{0.9,0.9,0.9}}c}{VSC} & \multicolumn{2}{c}{Wang's
   Ex} \\
   $n_K$ & $\gamma_{n_K}$ & $n_K$ & $\gamma_{n_K}$
   & $n_K$ & $\gamma_{n_K}$ & $n_K$ & $\gamma_{n_K}$\\
 & & & & & &\\
 \rowcolor[gray]{.7} $4*$ & $0.0737$ & $4*$ & $1.134$ & $4*$ & $3.216$ & $4$ & $50.640$ \\
 $4$ & $0.0924$ & $4$ & $1.134$ & $4$ & $3.421$ & $4$ & $50.640$\\
 $3$ & $0.0924$ & $3$ & $1.134$ & $3$ & $3.408$ & $3$ & $50.640$\\
 $2$ & $0.0925$ & $2$ & $1.134$ & $2$ & $3.437$ & $2$ & $50.642$\\
 $1$ & $0.1235$ & $1$ & $1.134$ & $1$ & $3.362$ & $1$ & $50.645$\\

 \rowcolor[gray]{1.0} & & & & & &\\
   \multicolumn{2}{>{\columncolor[rgb]{0.9,0.9,0.9}}c}{AUV Speed} & \multicolumn{2}{c}{AUV Heading}
   & \multicolumn{2}{>{\columncolor[rgb]{0.9,0.9,0.9}}c}{AUV Depth} &
   \multicolumn{2}{c}{PA} \\
   $n_K$ & $\gamma_{n_K}$ & $n_K$ & $\gamma_{n_K}$
   & $n_K$ & $\gamma_{n_K}$ & $n_K$ & $\gamma_{n_K}$\\
 & & & & & &\\
 \rowcolor[gray]{.7} $3$ & $0.954$ & $5$ & $0.954$ & $6$ & $0.955$ & $5$ & $1.0\;10^{-6}$\\
                     $3$ & $0.955$ & $5$ & $0.954$ & $6$ & $0.955$ & $5$ & $4.1\;10^{-4}$\\
                     $2$ & $0.955$ & $4$ & $0.954$ & $5$ & $0.955$ & $4$ & $3.6\;10^{-4}$\\
                     $1$ & $0.955$ & $3$ & $0.954$ & $4$ & $0.955$ & $3$ & $3.6\;10^{-4}$\\
                         &          & $2$ & $0.954$ & $3$ & $0.955$ & $2$ & $3.5\;10^{-4}$\\
                         &          & $1$ & $0.954$ & $2$ & $0.955$ & $1$ & $3.5\;10^{-4}$\\
                         &          &  &  & $1$ & $0.957$ &  & \\
 \rowcolor[gray]{1.0} & & & & & &\\
   \multicolumn{2}{>{\columncolor[rgb]{0.9,0.9,0.9}}c}{Enns' Ex} & \multicolumn{2}{c}{AC$8$}
   & \multicolumn{2}{c}{} & \multicolumn{2}{c}{} \\
   $n_K$ & $\gamma_{n_K}$ & $n_K$ & $\gamma_{n_K}$ & \multicolumn{2}{c}{}  \\
 & & & & \multicolumn{2}{c}{} \\
 \rowcolor[gray]{.7} $8$ & $1.127$ & $9$ & $1.617$ & \multicolumn{2}{c}{} \\
                     $8$ & $1.130$ & $9$ & $1.617$ & \multicolumn{2}{c}{} \\
                     $7$ & $1.131$ & $8$ & $1.617$ & \multicolumn{2}{c}{} \\
                     $6$ & $1.130$ & $7$ & $1.617$ & \multicolumn{2}{c}{} \\
                     $5$ & $1.131$ & $6$ & $1.617$ & \multicolumn{2}{c}{} \\
                     $4$ & $1.167$ & $5$ & $1.617$ & \multicolumn{2}{c}{} \\
                     $3$ & $1.200$ & $4$ & $1.617$ & \multicolumn{2}{c}{} \\
                     $2$ & $1.244$ & $3$ & $1.617$ & \multicolumn{2}{c}{} \\
                     $1$ & $1.426$ & $2$ & $1.626$ & \multicolumn{2}{c}{} \\
                         &          & $1$ & $1.651$ & \multicolumn{2}{c}{} \\
\multicolumn{8}{c}{} \\
 \cline{1-4}
\end{tabular}
\end{center}
\end{table}

Results obtained using \hifoo\ to find stable controllers for examples
$5-17$ are shown in Table \ref{otherexamples1_table} and
\ref{otherexamples2_table}. The examples are grouped according to
plant order: Table \ref{otherexamples1_table} shows low to
medium-order plants and Table \ref{otherexamples2_table} shows
higher-order plants. For the low to medium-order plants, we report
results for strong stabilization with order ranging from 1 to the
order of the plant. For the higher-order plants, we restricted the
order of the controller to $8$.


\begin{table}[h!]
\caption{Results of \hifoo\ on Other Benchmark Examples (High-Order
Plants). The first line shows the plant order and full-order
controller performance, with an asterisk indicating an unstable
full-order controller.} \label{otherexamples2_table} \vspace{-5mm}
\begin{center}
\begin{tabular}{>{\columncolor[rgb]{0.9,0.9,0.9}}c>{\columncolor[rgb]{0.9,0.9,0.9}}c
   cc
   >{\columncolor[rgb]{0.9,0.9,0.9}}c>{\columncolor[rgb]{0.9,0.9,0.9}}c
   cc}
   \hline
   \multicolumn{8}{c}{} \\
   \multicolumn{2}{>{\columncolor[rgb]{0.9,0.9,0.9}}c}{HIMAT} & \multicolumn{2}{c}{AC$10$}
   & \multicolumn{2}{>{\columncolor[rgb]{0.9,0.9,0.9}}c}{BDT$2$} & \multicolumn{2}{c}{HF$1$} \\
   $n_K$ & $\gamma_{n_K}$ & $n_K$ & $\gamma_{n_K}$
   & $n_K$ & $\gamma_{n_K}$ & $n_K$ & $\gamma_{n_K}$\\
 & & & & & &\\
 \rowcolor[gray]{.7} $20$ & $0.970$ & $55*$ & $0.633$ & $82*$ & $0.234$ & $130$ & $0.447$ \\
 $8$ & $1.060$ & $8$ & $8.984$ & $8$ & $0.531$ & $8$ & $0.447$ \\
 $7$ & $1.068$ & $7$ & $9.003$ & $7$ & $0.542$ & $7$ & $0.447$ \\
 $6$ & $1.061$ & $6$ & $9.376$ & $6$ & $0.534$ & $6$ & $0.447$ \\
 $5$ & $1.067$ & $5$ & $9.570$ & $5$ & $0.559$ & $5$ & $0.447$ \\
 $4$ & $1.072$ & $4$ & $9.869$ & $4$ & $0.604$ & $4$ & $0.447$ \\
 $3$ & $1.109$ & $3$ & $9.869$ & $3$ & $0.578$ & $3$ & $0.447$ \\
 $2$ & $2.155$ & $2$ & $9.869$ & $2$ & $0.576$ & $2$ & $0.447$ \\
 $1$ & $2.782$ & $1$ & $10.863$ & $1$ & $0.643$ & $1$ & $0.447$ \\
  \rowcolor[gray]{1.0} & & & & & &\\
   \multicolumn{2}{>{\columncolor[rgb]{0.9,0.9,0.9}}c}{CM$4$} & \multicolumn{2}{c}{}
   & \multicolumn{2}{c}{} & \multicolumn{2}{c}{} \\
   $n_K$ & $\gamma_{n_K}$ &  \multicolumn{4}{c}{}  \\
 & & & & \multicolumn{2}{c}{} \\
 \rowcolor[gray]{.7} $240$ & $0.816$ & \multicolumn{4}{c}{} \\
                     $8$ & $0.824$ & \multicolumn{4}{c}{} \\
                     $7$ & $0.819$ & \multicolumn{4}{c}{} \\
                     $6$ & $0.818$ & \multicolumn{4}{c}{} \\
                     $5$ & $0.817$ & \multicolumn{4}{c}{} \\
                     $4$ & $0.818$ & \multicolumn{4}{c}{} \\
                     $3$ & $0.817$ & \multicolumn{4}{c}{} \\
                     $2$ & $0.817$ & \multicolumn{4}{c}{} \\
                     $1$ & $0.817$ & \multicolumn{4}{c}{} \\
                         &         & \multicolumn{4}{c}{} \\
\multicolumn{8}{c}{} \\
 \cline{1-2}
\end{tabular}
\end{center}
\end{table}

The performance of \hifoo\ is very good for low and medium-order
plants. In most cases, the optimal closed-loop full-order
$\Hi$ performance is achieved or nearly achieved by
$1^{\textrm{st}}$-$3^{\textrm{rd}}$ order stable controllers, even
though the full-order controller is not necessarily stable. In
general, \hifoo\ shows that it is often possible to find a stable
low-order controller without greatly sacrificing closed-loop $\Hi$
performance.

\hifoo\ also performs successfully on higher-order plants as shown
in Table \ref{otherexamples2_table}. These examples are numerically
difficult and it is sometimes difficult to calculate the optimal
full-order $\Hi$ performance with well-known robust algorithms.
The plant AC10 is particularly difficult to stabilize so we needed
more runs than used for the other plants, building higher order
controllers with lower order ones as initial search points; we omit
the details.
These results clearly demonstrate that \hifoo\ is very effective not
only for simple plants but real-life high-order plants that arise in
industrial applications.


\section{Concluding Remarks} \label{concluding}
We applied the \hifoo\ Toolbox to $21$ strongly-stable design
problems, taken from a mix of industrial and academic test problems.
The performance of \hifoo\ is better than existing results in the
literature in most cases, even when the specified controller order
is low. We conclude that \hifoo\ is an effective method for
fixed-order stable $\Hi$ controller design, giving flexibility to
the designer to specify the controller order and generally obtaining
good performance. \hifoo, which is written in \matlab, is easy to use
and is freely available on the web.

\section{Acknowledgments}
The work of the second and third authors respectively was supported
in part by the U.S. National Science Foundation (NSF) Grant under
grants DMS-0602235 and DMS-0714321. The views expressed in this
paper are those of the authors and are not necessarily shared by the
NSF.


\begin{thebibliography}{99}
\bibitem{FourDiskSystem} D.~S.~Bernstein and W.~M.~Haddad, ``LQG
control with an $H_{\infty}$ performance bound: A riccati equation
approach,'' {\it IEEE Transactions on Automatic Control,} vol.34,
pp. 293--305, 1989.

\bibitem{BurkeTAC06} J.~V.~Burke, D.~Henrion, A.~S.~Lewis and
M.~L.~Overton, ``Stabilization via nonsmooth, nonconvex
optimization,'' {\it IEEE Transactions on Automatic Control,}
vol.51, pp. 1760--1769, 2006.

\bibitem{BHLO-IFAC-06} J. V. Burke, D. Henrion, A. S. Lewis, M. L.
Overton, ``HIFOO - A MATLAB package for fixed-order controller
design and $H_\infty$ optimization'', {\it IFAC Symposium on Robust
Control Design}, Toulouse, France, July 2006.

\bibitem{BurkeSIAM05} J.~V.~Burke, A.~S.~Lewis and M.~L.~Overton, ``A robust gradient sampling algorithm for
nonsmooth nonconvex optimization,'' {\it SIAM Journal on
Optimization,} vol.15, pp. 751--779, 2005.

\bibitem{BurkeOverton01} J.~V.~Burke and M.~L.~Overton,
``Variational Analysis of Non-Lipschitz Spectral Functions,''
{\it Mathematical Programming}, vol.90, pp. 317-352, 2001.

\bibitem{CZ-TAC-01}
D.~U.~Campos-Delgado and K.~Zhou,
  ``$\Hi$ Strong stabilization,'' {\it IEEE Transactions on
  Automatic Control,} vol.46, pp.1968-1972, 2001.

\bibitem{CZ-TAC-03}
D.~U.~Campos-Delgado and K.~Zhou, ``A parametric optimization
approach to $\Hi$ and $\mathcal{H}^2$ strong stabilizaton,'' {\it
Automatica}, vol.39, No.7, pp. 1205-1211, 2003.

\bibitem{CL-AUT-00}
Y. Y. Cao and J. Lam, ``On simultaneous $H_\infty$ control and
strong $H_\infty$ stabilization,'' {\it Automatica}, vol.36, No.6,
pp. 859-–865, 2000.

\bibitem{Piezoelectric} B.~M.~Chen, $H_\infty$ Control and Its Applications, vol. 235 of Lectures
Notes in Control and Information Sciences, Springer Verlag, New
York, Heidelberg, Berlin, 1998.

\bibitem{CC-TAC-01}
Y. Choi and W.K. Chung, ``On the Stable $\Hi$ Controller
Parameterization Under Sufficient Condition,'' {\it IEEE
Transactions on Automatic Control,} vol.46, pp. 1618-1623, 2001.

\bibitem{CLC-IET-07}
Y.~S. Choi, J.~L. Leu and Y.~C.~Chung, ``Stable controller design
for MIMO systems: an LMI approach,'' {\it IET Control Theory
Applications,} vol.1, pp. 817-829, 2007.

\bibitem{CWL-CDC-03}
Y.~S. Chou, T. Z. Wu and J. L. Leu, ``On Strong Stabilization and
$\Hi$ Strong-Stabilization Problems,'' {\it Proc. Conference on
Decision and Control}, pp. 5155--5160, 2003.

\bibitem{AC10example} E.~J.~Davison, ``Benchmark problems for
control system design,'' technical report, International Federation
of Automatic Control, 1990. Report of the IFAC Theory Comittee.

\bibitem{DGKF} J.~Doyle, K.~Glover, P.~Khargonekar, and B.~A.~Francis, ``State-space solutions
to standard $\mathcal{H}_2$ and $\mathcal{H}_\infty$ control
problems,''
 {\it IEEE Transactions on Automatic Control,} vol. 34, no. 8, pp. 831–-847, 1989.

\bibitem{Enns_example} D.F.~Enns, ``Model Reduction for Control System Design,'' Ph.D. Dissertation, Stanford
University, 1984.

\bibitem{AUVEx} Z.~Feng and R.~Allen, ``Reduced order $H_\infty$
control of an autonomous underwater vehicle,'' {\it Control
Engineering Practice,} vol.12, pp. 1511--1520, 2004.

\bibitem{RobustControlToolbox}
G.~J.~Galas, A.~K.~Packard, M.~G.~Safonov and R.~Y.~Chiang, {\it
Robust Control Toolbox $3.3$,} The Mathworks, Natick, 2008.

\bibitem{TransportPlane} D.~Gangsaas, K.~Bruce, J.~Blight, and U.-L.~Ly, ``Application of
modern synthesis to aircraft control: Three case studies,'' {\it
IEEE Transactions on Automatic Control,} vol.31, pp. 995--1014,
1986.

\bibitem{Wang2006} H.~Gao, J.~Lam and C.~Wang, ``Controller reduction with $H_\infty$
error performance: continuous- and discrete-time cases'', {\it
International Journal of Control,} vol.79, pp. 604--616, 2006.

\bibitem{HIMATExGod} P.~J.~Goddard and K.~Glover, ``Controller Approximation: Approaches for preserving
$H_\infty$ Performance,'' {\it IEEE Transactions on Automatic
Control,} vol.36, pp. 858--871, 1998.

\bibitem{HIFOO_08} S.~Gumussoy and M.~L.~Overton, ``Fixed-Order H-infinity Controller
Design via HIFOO, a Specialized Nonsmooth Optimization Package,'',
{\it Proc. of American Control Conference,} pp. 2750--2754, 2008.

\bibitem{HinfssMIMO_GO} S.~Gumussoy and H.~\"Ozbay, ``Remarks on strong stabilization and stable
$\Hi$ controller design,'' {\it IEEE Transactions on Automatic
Control,} vol.50, pp. 2083--2087, 2005.

\bibitem{HinrPrit05} D.~Hinrichsen and A.~J.~Pritchard,
Mathematical Systems Theory I: Modelling, State Space Analysis,
Stability and Robustness, Springer, 2005.

\bibitem{HFexample} A.~S.~Hodel, K.~Poolla, and B.~Tension,
``Numerical solution of the Lyapunov equation by approximate power
iteration,'' {\it Linear Algebra Applications,} vol.236, pp.
205--230, 1996.

\bibitem{Chemicalreactor} Y.~S.~Hung and A.~G.~J.~MacFarlane, Multivariable feedback: A
classical approach, Lectures Notes in Control and Information
Sciences, Springer Verlag, New York, Heidelberg, Berlin, 1982.

\bibitem{Enns_ex2001} D.~Kavranoglu and S.H.~Al-Amer, ``New efficient frequency domain algorithm for
$\Hi$ approximation with applications to controller reduction,''
{\it IEE Proceedings-Control Theory Applications,} vol.148, pp.
383--390, 2001.

\bibitem{VTOL} L.~H.~Keel, S.~P.~Bhattacharyya, and J.~W.~Howze, ``Robust control
with structured perturbations,'' {\it IEEE Transactions on Automatic
Control,} vol.36, pp. 68--77, 1988.

\bibitem{Compleib} F. Leibfritz, ``COMPL$_{e}$IB, constraint matrix-optimization problem
library— A collection of test examples for nonlinear semidefinite
programs, control system design and related problems,''
Universit\"at Trier, Tech. Rep., 2003. www.compleib.de

\bibitem{LS-SCL-02}
P.H.~Lee and Y.C.~Soh,
  ``Synthesis of stable $\Hi$ controller via the chain scattering framework,''
  {\it System and Control Letters,} vol.46, pp.1968--1972, 2002.

\bibitem{VSC1997} J.~S.~Lin, and I.~Kanellakopoulos, ``Nonlinear design of active
suspensions,'' {\it IEEE Control Systems Magazine,} vol.17, pp. 45–-59, 1997.

\bibitem{Millstone2006} M.~Millstone,
``HIFOO 1.5: Structured control of linear systems with a non-trivial
feedthrough", M.S.\ thesis, Courant Institute of Mathematical
Sciences, New York University, 2006.

\bibitem{CMexample} A.~H.~Nayfeh, J.~F.~Nayfeh, and D.~T.~Mook, ``On methods for
continuous systems with quadratic and cubic nonlinearities,'' {\it
Nonlinear Dynamics,} vol.3, pp. 145--162, 1992.

\bibitem{BDTexample} S.~Skogestad and I.~Postlethwaite,
Multivariable Feedback Control, John Wiley \& Sons, 1996.

\bibitem{SmithSongerald86} M.~C. Smith and K.~P. Sondergeld, ``On the order of stable
compensators,'' {\it Automatica,} vol.22, pp. 127--129, 1986.

\bibitem{V-CSS-85}
 M.~Vidyasagar,
 {\it Control System Synthesis: A Factorization
  Approach}, MIT Press, 1985.

\bibitem{Wang2003} J.-Z.~Wang and L.~Huang, ``Controller order
reduction with guaranteed performance via coprime factorization'',
{\it International Journal of Robust and Nonlinear Control,} vol.13,
pp. 501--517, 2003.

\bibitem{ZO-TAC-99} M.~Zeren and H.~\"Ozbay,
``On the synthesis of stable $\Hi$
  controllers,'' {\it IEEE Transactions on
  Automatic Control,} vol.44, pp. 431--435, 1999.

\bibitem{ZerenOzbayAutomatica} M.~Zeren and H.~\"Ozbay,
``On the strong stabilization and stable $H^\infty$-controller
design problems for MIMO systems,'' {\it Automatica,} vol.36, pp.
1675--1684, 2000.

\bibitem{ZDG-ROC-96}
K.~Zhou, J.C.~Doyle and K.~Glover,
 {\it Robust and Optimal Control}, Upper Saddle River: Prentice-Hall, 1996.
\end{thebibliography}
\end{document}